\documentclass[12pt]{article}
\begin{document}

\addtolength{\baselineskip}{0.5\baselineskip}

\title{\textbf{Second Quantized Reduced Bloch Equations and the
Exact Solutions for Pairing Hamiltonian}}
\author{Liqiang Wei \\
Institute for Theoretical Atomic, Molecular and Optical Physics\\
Harvard University, Cambridge, MA 02138\\
\\
Chiachung Sun\\
Institute of Theoretical Chemistry, Jilin University\\
Changchun, Jilin 130023 P. R. China}

\maketitle

\begin{abstract}
\vspace{0.05in}

In this article, we present a set of hierarchy Bloch equations for
the reduced density operators in either canonical or grand
canonical ensembles in the occupation number representation. They
provide a convenient tool for studying the equilibrium quantum
statistical mechanics for some model systems. As an example of
their applications, we solve the equations for the model system
with a pairing Hamiltonian. With the aid of its symplectic group
symmetry, we obtain the statistical reduced density matrices with
different orders. As a special instance for the solutions, we also
get the reduced density matrices of the ground state for a
superconductor.
\end{abstract}

\vspace{0.35in}
\section{Introduction}

The language of the second quantization is a convenient and
ubiquitous tool for studying the quantum many-body systems in
modern physics. The quantum statistics of the identical many
particles is automatically satisfied by the commutation relations
of the creation or annihilation operators, and the complicated
mathematical manipulation with wavefunctions such as Slater
determinants and with matrix elements of operators is
significantly simplified. It is no surprise that a familiarity
with the related knowledge of the occupation number representation
is a must for a serious theorist.

In a recent paper, we have developed an equation of motion
approach for a direct determination of the reduced density
matrices for the statistical ensembles under the framework of
equilibrium quantum statistical mechanics [1-4]. We have derived
the hierarchy Bloch equations which provide the laws according to
which the reduced statistical density operators vary in
temperature. We also take the orbital approximation, and solve the
equations for the identical fermion systems with two-body
interaction in a grand canonical ensemble. We have gained
significant new physical insights. We find that the usual
Fermi-Dirac statistics for the free fermion gas holds also for the
interacting fermions, and that the usual Hartree-Fock equation at
zero-temperature can be extended to the case of any finite
temperature. The mean force field that every orbital or particle
is subject to will be from the contribution of both coherent and
incoherent superpositions of the wavefunctions of other orbitals
or particles, and is therefore temperature-dependent [5-8]. This
opens a way for investigating the interplay between the
microscopic structure and macroscopic thermodynamic quantities for
molecules and solids [5-9]. In addition, the Fermi-Dirac
distribution for interacting particles demonstrates that there
exists an effective one-particle description for the interacting
many-body system [2,10].

In the current paper, we begin the presentation of the scheme for
an $\it{exact}$ solution of the reduced Bloch equations which will
be developed in a second quantized form. In quantum many-body
theory, the systems are often represented by the model
Hamiltonians in an occupation number representation. We study the
pairing Hamiltonian, which is a useful model for the study of
nuclear structure [11,12], microscopic state of superconductors
[13,14], and electronic structure of molecules [15,16].

We organize this paper as follows. In the next Section, starting
from the Bloch equation, we deduce a set of hierarchy equations in
the occupation number representation for the reduced density
operators in the canonical ensemble. The derivation is based on
the second quantized form of the contraction operator which is
presented in the Appendix. We also extend it to the case of a
grand canonical ensemble. In Section 3, we first briefly review
the symplectic group symmetry of the pairing Hamiltonian and
present the second quantized form of its eigenvectors
characterized by the particle number, the weight and the seniority
number. Then using the symmetric conditions among $D^{1}$, $D^{2}$
and $D^{3}$, we solve the reduced Bloch equations in the canonical
ensemble for the pairing Hamiltonian and obtain its canonical
reduced density matrices with different orders. As a special
situation, they yield the reduced density matrices of the ground
state for the superconductor.  We conclude our discussion in the
final Section.

\vspace{0.35in}

\section{Reduced Bloch Equations in the Occupation Number
Representation}

We consider an identical $N$-fermion system with a two-body
interaction. Its Hamiltonian has the following form,
\begin{equation}
 H = \sum_{i=1}^{N}h(i) + \sum_{i<j}^{N} v(i,j),
\end{equation}
where $h(i)$ and $v(i,j)$ are one- and two-particle operators,
respectively. Assume a generic and complete set of single-particle
states $\{|i>\}$, which form a single-particle Hilbert space $V$.
For each $|i>$, the corresponding creation or annihilation
operators are defined as
\begin{equation}
  |i> = a^{+}_{i}|0>,  \   a_{i}|j> = \delta_{ij}|0>.
\end{equation}
They satisfy the anticommutation relations,
\begin{eqnarray}
\nonumber \{a_{i}, a^{+}_{j}\} &=& \delta_{ij}, \\
 \{a_{i}, a_{j}\} = \{a^{+}_{i}, a^{+}_{j}\} &=& 0.
\end{eqnarray}
Define an $N$-fold antisymmetrized direct product space of the
single-particle Hilbert space $V$ as
\begin{equation}
V^{N\wedge} = V_{1}\wedge V_{2}\wedge ... \wedge V_{N}.
\end{equation}
The Fock space $F$ can be written as
\begin{equation}
F = \sum_{N=0}^{\infty} \oplus V^{N\wedge},
\end{equation}
and any operator $T$ on $F$ can be expressed as the direct sum of
the $N$-fold tensor $T^{N}$,
\begin{equation}
T = \sum_{N=0}^{\infty} \oplus T^{N}.
\end{equation}
The contraction operator, which defines the reduced density matrix
as follows [1,17,18]
\begin{equation}
  D^{p} = L_{N}^{p} (D^{N}),
\end{equation}
is a superoperator acting on the $N$th-order density matrix
$D^{N}$, and has following second quantized form (see the
Appendix),
\begin{equation}
  L_{N}^{p}(T^{N}) = \frac{p!}{N!} (\Lambda_{-})^{N-p}(T^{N}),
\end{equation}
where $\Lambda_{-}$ is defined by
\begin{equation}
\Lambda_{-}(T^{N}) = \sum_{i}a_{i}(T^{N})a^{+}_{i}.
\end{equation}

Equipped with these definitions as well as the Hamiltonian in the
occupation number representation,
\begin{equation}
  H = \sum_{ij}h_{ij}a_{i}^{+}a_{j} +
  \frac{1}{2}\sum_{ijkl}v_{ijkl}a^{+}_{i}a^{+}_{j}a_{l}a_{k},
\end{equation}
where $h_{ij}=<i|h|j>$ and $v_{ijkl}=<ij|v|kl>$, we can now deduce
the second quantized reduced Bloch equations [19-25]. For the
canonical ensemble, the Bloch equation is [1-4]
\begin{equation}
-\frac{\partial}{\partial \beta}D^{N} =
\left(\sum_{ij}h_{ij}a_{i}^{+}a_{j} +
  \frac{1}{2}\sum_{ijkl}v_{ijkl}a^{+}_{i}a^{+}_{j}a_{l}a_{k}\right)D^{N},
\end{equation}
where the $D^{N}$ is regarded as an $N$th-order irreducible
tensor, and $\beta$ is the reverse of the product of the Boltzmann
constant $k_{B}$ and absolute temperature $T$. Applying on both
sides of above equation with the operator $L_{N}^{p}$ gives
\begin{equation}
-\frac{\partial}{\partial \beta}D^{p} =
\sum_{ij}h_{ij}L_{N}^{p}\left(a_{i}^{+}a_{j}D^{N}\right) +
  \frac{1}{2}\sum_{ijkl}v_{ijkl}L_{N}^{p}\left(a^{+}_{i}a^{+}_{j}a_{l}a_{k}D^{N}\right).
\end{equation}
The evaluation of the first term of the right hand side of the Eq.
(12) results in
\begin{equation}
\frac{N!}{p!}L_{N}^{p}\left(a^{+}_{i}a_{j}D^{N}\right)=(N-p)\frac{N!}{(p+1)!}a_{j}
L_{N}^{p+1}\left(D^{N}\right)a^{+}_{i}+
\frac{N!}{p!}a^{+}_{i}a_{j}L_{N}^{p}\left(D^{N}\right),
\end{equation}
and the calculation of the second term yields
\begin{eqnarray}
\nonumber\frac{N!}{p!}L_{N}^{p}\left(a^{+}_{i}a^{+}_{j}a_{l}a_{k}D^{N}\right)&=&\frac{(N-p)(N-p-1)N!}
{(p+2)!}a_{l}a_{k}L_{N}^{p+2}(D^{N})a^{+}_{i}a^{+}_{j}+\\
\nonumber&&+\frac{(N-p)N!}{(p+1)!}\left\{a^{+}_{j}a_{l}a_{k}L^{p+1}_{N}(D^{N})a^{+}_{i}-
a^{+}_{i}a_{l}a_{k}L^{p+1}_{N}(D^{N})a^{+}_{j}\right\}+ \\
&&+\frac{N!}{p!}a^{+}_{i}a^{+}_{j}a_{l}a_{k}L_{N}^{p}(D^{N}).
\end{eqnarray}
Therefore, the set of reduced Bloch equations for the canonical
ensemble in the occupation number representation is
\begin{eqnarray}
\nonumber
-\frac{\partial}{\partial\beta}D^{p}=HD^{p}+\frac{N-p}{p+1}\left(\sum_{ij}h_{ij}a_{j}
D^{p+1}a^{+}_{i}+\frac{1}{2}\sum_{ijkl}v_{ijkl}a^{+}_{j}a_{l}a_{k}D^{p+1}a^{+}_{i}-\right.\\
\left.-\frac{1}{2}\sum_{ijkl}v_{ijkl}a^{+}_{i}a_{l}a_{k}D^{p+1}a^{+}_{j}\right)
+\frac{(N-p)(N-p-1)}{2(p+2)(p+1)}\sum_{ijkl}v_{ijkl}a_{l}a_{k}D^{p+2}a^{+}_{i}a^{+}_{j}.
\end{eqnarray}
Define
\begin{equation}
    \hat{v}_{ijkl} = \frac{1}{2}(v_{ijkl}-v_{jikl}),
\end{equation}
then the above equation can be simplified as
\begin{eqnarray}
\nonumber
-\frac{\partial}{\partial\beta}D^{p}&=&HD^{p}+\frac{N-p}{p+1}\left(\sum_{ij}h_{ij}a_{j}
D^{p+1}a^{+}_{i}+\sum_{ijkl}\hat{v}_{ijkl}a^{+}_{j}a_{l}a_{k}D^{p+1}a^{+}_{i}\right)+\\
&&+\frac{(N-p)(N-p-1)}{2(p+2)(p+1)}\sum_{ijkl}\hat{v}_{ijkl}a_{l}a_{k}D^{p+2}a^{+}_{i}a^{+}_{j}.
\end{eqnarray}

For the grand canonical ensemble, whose $p$th-order reduced
density matrix is defined by [1]
\begin{equation}
 D^{p}_{G} =\sum_{N=p}^{\infty}\oplus
 \left(\begin{array}{c}N\\p\end{array}\right)L_{N}^{p}[D^{N}],
\end{equation}
similar procedure leads to the following equation
\begin{eqnarray}
\nonumber
-\frac{\partial}{\partial\beta}D^{p}_{G}&=&\bar{H}D^{p}_{G}+\sum_{ij}\bar{h}_{ij}
a_{j}D^{p+1}_{G}a^{+}_{i}+\sum_{ijkl}\hat{v}_{ijkl}a^{+}_{j}a_{l}a_{k}D^{p+1}_{G}a^{+}_{i}+\\
&&+\frac{1}{2}\sum_{ijkl}\hat{v}_{ijkl}a_{l}a_{k}D^{p+2}_{G}a^{+}_{i}a^{+}_{j},
\end{eqnarray}
where
\begin{equation}
  \bar{h}_{ij}=h_{ij}-\mu \delta_{ij},
\end{equation}
and
\begin{equation}
  \bar{H} = \sum_{ij}\bar{h}_{ij}a_{i}^{+}a_{j} +
  \frac{1}{2}\sum_{ijkl}v_{ijkl}a^{+}_{i}a^{+}_{j}a_{l}a_{k}.
\end{equation}
This is a set of second quantized reduced Bloch equations for the
reduced density matrices in the grand canonical ensemble.

\vspace{0.35in}

\section{Reduced Density Matrices with Different Orders for
Pairing Hamiltonian}

\subsection{The Pairing Hamiltonian and its Eigenvalues and
Eigenstates}

Prior to solve the reduced Bloch equations in the case of the
canonical ensemble for a pairing Hamiltonian, we need to review
its eigenvalue problem as well as related symmetry properties
which is discussed in details in paper [26].

Label a set of $2\lambda$ spin-orbitals $|\psi_{i}>$ $(dim\
 V=2\lambda=r)$ as
\[
  1,...,i,...,\lambda, 1^{'},...,i^{'},...,\lambda^{'},
\]
where $i^{'}=i+\lambda$, and $|i^{'}>$ represents the pairing of a
single-particle state $|i>$. The pairing Hamiltonian is defined as
[26-29]
\begin{eqnarray}
\nonumber H &=& \epsilon N- GQ_{+}Q_{-} \\
&=&\epsilon\sum_{i=1}^{r}a_{i}^{+}a_{i}-G\sum_{i,j=1}^{\lambda}a_{i}^{+}a_{i^{'}}^{+}
a_{j^{'}}a_{j},
\end{eqnarray}
where
\[   Q_{+} = \sum_{i=1}^{\lambda}b_{i}^{+},\
b_{i}^{+}=a_{i}^{+}a^{+}_{i^{'}},  \] and
\begin{equation}
Q_{-} = \sum_{i=1}^{\lambda}b_{i},\  b_{i}=a_{i^{'}}a_{i},
\end{equation}
are called the quasi-spin operators. They are the generators of
the Lie algebra for the quasi-spin group $SU^{Q}(2)$. The second
term of the Hamiltonian indicates that the interaction is only for
the pairing states of the particles, and is therefore called the
pairing force. The matrix element of the interaction is
represented by a constant $G$. When $G>0$, the interaction is
repulsive; when $G<0$, the interaction is attractive.

Since the pairing Hamiltonian Eq. (22) exhibits a symplectic
symmetry, its $N$-particle eigenstates should be the basis vectors
of the irreducible representation space of the symplectic group,
and can be characterized by its weight [30]. For the symplectic
group $Sp(2\lambda)$ and its corresponding Lie algebra
$C_{\lambda}$, the weight is the eigenvalue for the generator,
\[  S_{ii} = a^{+}_{i}a_{i}-a^{+}_{i^{'}}a_{i^{'}},\  i
=1,2,...,\lambda, \] which is the difference of occupation numbers
between the pairing spin orbitals $|i>$ and $|i^{'}>$, with the
value being $1$, $0$, or $-1$. They can be grouped together as a
vector
\[ (\vec{S})=(s_{1},s_{2}, ...,s_{\lambda}).\]
 Furthermore, every
irreducible representation can be characterized by the maximal
weight. For the symplectic group $Sp(2\lambda)$, this maximal
weight takes the form of $(1^{\nu}\ 0^{\lambda-\nu})$, where the
number $\nu$ is called the seniority number, and the irreducible
representation of the symplectic group can be written as
$<1^{\nu}>$. However, there is a possibility that the same weight
might still correspond to the different basis vectors. To
completely distinguish all the basis vectors of an irreducible
representation space for the symplectic group, we therefore need
to classify the $N$-particle eigenstates according to the
following group chain
\begin{equation}
 Sp(2\lambda)\supset Sp(2\lambda-2)\supset ...\supset
Sp(2).
\end{equation}
This means that each basis vector will further be characterized by
a group of $\lambda$ seniority numbers: \[ (\vec{V})=(\nu_{1}\
\nu_{2}\ ...\ \nu_{\lambda}). \] Thus, the eigenvectors of the
pairing Hamiltonian will be completely identified by the particle
number $N$, the weight $(\vec{S})$, and seniority number
$(\vec{V})$: $|N,(\vec{S}),(\vec{V})>$. They belong to the
irreducible representation spaces $<1^{\nu_{1}}>$,
$<1^{\nu_{2}}>$,..., and $<1^{\nu_{\lambda}}>$ of the group
$Sp(2)$, $Sp(4)$, ..., and $Sp(2\lambda)$, respectively, and
satisfy the eigenequation
\begin{equation}
S_{ii}|N,(\vec{S}),(\vec{V})> = s_{i}|N,(\vec{S}),(\vec{V})>,\
(s_{i}=1,0,-1).
\end{equation}
The branching rule of $V^{\wedge\Lambda}$ under the group chain
(24) of $S_{p}(2\lambda)$ is
\[ [1^{N}] \rightarrow <1^{N}> +<1^{N-2}> + ...+ <1> or <0>, \]
 for $N\le \lambda$, or
 \begin{equation}
 [1^{N}] \rightarrow <1^{2\lambda-N}>+<1^{2\lambda-N-2}> + ...+ <1> or <0>
 \end{equation}
 for $N> \lambda$. The dimension of the irreducible space $<1^{\nu}>$ is given by
\begin{equation}
  dim \ <1^{\nu}> = \left(\begin{array}{c}
  2\lambda\\ \nu \end{array}\right)-\left(\begin{array}{c}
  2\lambda\\ \nu-2 \end{array}\right).
\end{equation}
The analytical expression for the basis vectors can be constructed
as the following form [26]
\begin{equation}
|N,(\vec{S}),(\vec{V})> = |N, <1^{\nu}>,W> =
\sqrt{\frac{(\lambda-\nu-m)!}{m!(\lambda-\nu)!}}Q^{m}_{+}g^{+}_{k_{1}}
...g^{+}_{k_{y}}a^{+}_{j_{1}}...a^{+}_{j_{x}}|0>,
\end{equation}
where
\begin{equation}
g^{+}_{k}=\frac{1}{\sqrt{(k-\nu_{k}+2)(k-\nu_{k}+1)}}\left[\sum_{i=1}^{k-1}b^{+}_{i}
-(k-\nu_{k}+1)b^{+}_{k}\right],
\end{equation}
and \[ m = \frac{1}{2}(N-\nu).\] In Eq. (28), the $k_{i}$ is the
positive number which gives the difference of two consecutive
seniority numbers in the $(\vec{V})$ which is two:
$\nu_{k}-\nu_{k-1}=2$; the $j_{i}$ means the same thing but its
value equals one: $\nu_{|j|}-\nu_{|j|-1}=1$. The sign of $j_{i}$
is decided by the sign of $s_{|j|}$: if $s_{|j|} =1,\ j > 0$;
while if $s_{|j|} =-1,\ j < 0$. In other words, $\pm k_{1}, ...,
\pm k_{y}$ are the pairing numbers in the corresponding symplectic
table $W$, while $j_{1}, ..., j_{x}$ are the unpairing numbers
appearing in the table, which captures all the features of the
basis vectors (28) classified according to the group chain (24)
[26,31,32].

The actions of the quasi-spin operators $Q_{+}$ or $Q_{-}$ on the
eigenvectors are
\begin{equation}
 Q_{+}|N,(\vec{S}),(\vec{V})>=\frac{1}{2}\sqrt{(N-\nu+2)(r-N-\nu)}|N+2,(\vec{S}),(\vec{V})>,
\end{equation}
or
\begin{equation}
 Q_{-}|N,(\vec{S}),(\vec{V})>=\frac{1}{2}\sqrt{(N-\nu)(r-N-\nu+2)}|N-2,(\vec{S}),(\vec{V})>,
\end{equation}
respectively. Therefore, we can get the analytic expression for
the energy eigenvalue as follows
\begin{equation}
 E(\nu) = N\epsilon - \frac{G}{4}(N-\nu)(r-N-\nu+2).
\end{equation}

\vspace{0.15in}
\subsection{The Exact Solutions of Reduced Bloch Equation}

We are now in a position to solve the reduced Bloch equation (15)
for the pairing Hamiltonian (22). For the case of $p=1$, the set
of hierarchy equations can be written as
\begin{eqnarray}
\nonumber -\frac{\partial}{\partial \beta} D^{1}&=&N\epsilon D^{1}
-\frac{1}{2}(N-1)G\sum_{i}a^{+}_{i^{'}}Q_{-}D^{2}a^{+}_{i}+\frac{1}{2}(N-1)G\sum_{i}a^{+}_{i}Q_{-}D^{2}a^{+}_{i^{'}}  \\
&&- \frac{(N-1)(N-2)}{6}GQ_{-}D^{3}Q_{+}.
\end{eqnarray}
Since the pairing Hamiltonian is invariant under the symplectic
group $Sp(2\lambda)$, the density matrix describing the states of
the system can be characterized by its irreducible representations
as follows
 \begin{equation}
  D^{N}(\beta) = \sum_{\nu} e^{-\beta E(\nu)} D^{N}(\nu),
\end{equation}
 where
 \begin{equation}
  D^{N}(\nu)=\frac{1}{dim<1^{\nu}>} \sum_{W\in<1^{\nu}>}
  |N,<1^{\nu}>,W><N,<1^{\nu}>,W|
\end{equation}
and
 \begin{equation}
   Tr D^{N}(\nu) = 1\ \ (\nu=N,N-2,...,1 \ or\ 0).
\end{equation}
 In addition,
  \begin{equation}
   \sum_{\nu} e^{-\beta E(\nu)} = Z(\beta,\nu,N).
\end{equation}
 For the same reason, the $p$th-order reduced density matrix of
$N$th-order density matrix can be written as
\begin{equation}
  D^{p}(\beta) =
  \sum_{\nu,\nu^{'}}\omega_{p}(\nu,\nu^{'},\beta)D^{p}(\nu^{'}),
\end{equation}
where
 \begin{equation}
  D^{p}(\nu^{'})=\sum_{W\in
  <1^{\nu^{'}}>}|p,<1^{\nu^{'}}>,W><p,<1^{\nu^{'}}>,W|
\end{equation}
\[ (\nu^{'}=p,p-2,...,1 \ or\  0) \]
and
\begin{equation}
  Tr D^{p}(\beta) = Z(\beta, V, N).
\end{equation}
The thermal probability coefficients
$\omega_{p}(\nu,\nu^{'},\beta)$ are to be decided.

According to the branching rule (26) and dimension formula (27),
the one-particle Hilbert space $V$ (denoted as $<1>$) is
irreducible and its dimension is $r$. Thus
\begin{equation}
  D^{1}(\beta) = \sum_{\nu}\omega_{1}(\nu,1,\beta) D^{1}(1).
\end{equation}
The two-particle Hilbert Space $V^{2\wedge}$ comprises two
irreducible subspaces $<0>$ and $<1^{2}>$ of $Sp(2\lambda)$. Their
dimensions are $1$ and $\left( \begin{array}{c}
r\\2\end{array}\right)-1$, respectively. Therefore
\begin{equation}
  D^{2}(\beta)=\sum_{\nu}\omega_{2}(\nu,0,\beta)D^{2}(0)+\sum_{\nu}\omega_{2}(\nu,
  2,\beta)D^{2}(2).
\end{equation}
The three-particle Hilbert space $V^{3\wedge}$ also includes two
irreducible subspaces $<1>$ and $<1^{3}>$ of $Sp(2\lambda)$ with
dimensions being $r$ and $\left( \begin{array}{c}
r\\3\end{array}\right)-\left( \begin{array}{c}
r\\1\end{array}\right) $, respectively. Thus
\begin{equation}
  D^{3}(\beta)=\sum_{\nu}\omega_{3}(\nu,1,\beta)D^{3}(1)+\sum_{\nu}\omega_{3}(\nu,
  3,\beta)D^{3}(3).
\end{equation}
The substitution of the expressions (41), (42), and (43) into Eq.
(33) then leads to an equation that the above five coefficients
satisfy. Together with three trace conditions (40) of $D^{1}$,
$D^{2}$ and $D^{3}$ as well as the $N$-representability conditions
among $D^{1}$, $D^{2}$ and $D^{3}$ as shown below
\begin{eqnarray}
\nonumber D^{2}(\beta)&=& L^{2}_{3}\left[D^{3}(\beta)\right] \\
  &=& \sum_{\nu}\omega_{3}(\nu,1,\beta)L^{2}_{3}\left[D^{3}(1)\right]+\sum_{\nu}\omega_{3}(\nu,
  3,\beta)L^{2}_{3}\left[D^{3}(3)\right],
\end{eqnarray}
and
\begin{eqnarray}
\nonumber D^{1}(\beta)&=& L^{1}_{3}\left[D^{3}(\beta)\right] \\
  &=& \sum_{\nu}\omega_{3}(\nu,1,\beta)L^{1}_{3}\left[D^{3}(1)\right]+\sum_{\nu}\omega_{3}(\nu,
  3,\beta)L^{1}_{3}\left[D^{3}(3)\right],
\end{eqnarray}
it seems that we can completely determine the reduced density
matrices for the pairing Hamiltonian. However, due to the
complexity of the $N$-representability, it is often replaced by
other conditions. Here, we choose the symmetry conditions among
$D^{1}$, $D^{2}$ and $D^{3}$.

Suppose
\begin{equation}
 L^{2}_{3}\left[D^{3}(1)\right] = a_{1}D^{2}(0) +b_{1}D^{2}(2).
\end{equation}
From
\begin{equation}
  Q_{+}Q_{-}=\lambda\left(\begin{array}{c}N\\2\end{array}\right)\Gamma^{N}_{2}(|g><g|),
\end{equation}
where $\Gamma^{N}_{2}$ is the expansion operator (see references
[17,31] or the Appendix), we get
\begin{eqnarray}
\nonumber a_{1}&=& Tr\left[|g><g|L^{2}_{3}(D^{3}(1))\right] \\
  &=& \frac{1}{3}(r-2).
\end{eqnarray}
But
\begin{equation}
TrL^{2}_{3}\left[D^{3}(1)\right]=a_{1}+b_{1}\frac{1}{2}(r+1)(r-2)=r,
\end{equation}
we therefore have
\begin{equation}
 b_{1}=\frac{4}{3(r-2)}.
\end{equation}
In the same manner, we assume
\begin{equation}
 L^{2}_{3}\left[D^{3}(3)\right] = a_{2}D^{2}(0) +b_{2}D^{2}(2),
\end{equation}
and obtain
\begin{equation}
 a_{2} = 0,\   b_{2}= \frac{r(r-4)}{3(r-2)}.
\end{equation}
The insertion of Eqs. (46) and (51) into Eq. (44) yields
\begin{equation}
D^{2}(\beta)=\sum_{\nu}\omega_{3}(\nu,1,\beta)\frac{r-2}{3}D^{2}(0)+\sum_{\nu}\left[\omega_{3}(\nu,1,\beta)
\frac{4}{3(r-2)}+\omega_{3}(\nu,3,\beta)\frac{r(r-4)}{3(r-2)}\right]D^{2}(2).
\end{equation}
Similarly,
\begin{eqnarray}
\nonumber L^{1}_{3}(D^{3}(1)) &=& D^{1}(1), \\
 L^{1}_{3}(D^{3}(3)) &=& \frac{1}{6}(r+1)(r-4)D^{1}(1),
\end{eqnarray}
 and we get the expression for $D^{1}(\beta)$
 \begin{equation}
D^{1}(\beta)=\sum_{\nu}\left[\omega_{3}(\nu,1,\beta)+\omega_{3}(\nu,3,\beta)
\frac{1}{6}(r+1)(r-4)\right]D^{1}(1).
\end{equation}
Therefore, with the aid of symmetry conditions, we have expressed
the thermal probability coefficients of $D^{1}(\beta)$ and
$D^{2}(\beta)$ by those of $D^{3}(\beta)$.

Now we can calculate three terms of the right hand side of Eq.
(33) as follows,
\begin{eqnarray}
\sum_{i}a^{+}_{i^{'}}Q_{-}D^{2}(\beta)a^{+}_{i} &=& \sum_{\nu}\omega_{3}(\nu,1,\beta)
\frac{r-2}{3}\sum_{i=1}^{\lambda}|i^{'}><i^{'}|, \\
 \sum_{i}a^{+}_{i}Q_{-}D^{2}(\beta)a^{+}_{i^{'}} &=& -\sum_{\nu}\omega_{3}(\nu,1,\beta)
\frac{r-2}{3}\sum_{i=1}^{\lambda}|i><i|, \\
Q_{-}D^{3}(\beta)Q_{+} &=&\sum_{\nu}\omega_{3}(\nu,1,\beta)
\frac{r-2}{2}\sum_{i=1}^{r}|i><i|.
\end{eqnarray}
Assume
\begin{eqnarray}
\nonumber \omega_{3}(\nu,1,\beta) &=& e^{-\beta E(\nu)}\omega_{3}(\nu,1), \\
\omega_{3}(\nu,3,\beta) &=& e^{-\beta E(\nu)}\omega_{3}(\nu,3),
\end{eqnarray}
where
\[    exp(-\beta E(\nu)), \ \nu=N, N-2, ..., 1, \]
are linearly independent. Then from the trace condition of
$D^{3}(\beta)$,
\begin{equation}
Tr D^{3}(\beta)=\sum_{\nu}exp(-\beta
E(\nu))\left[\omega_{3}(\nu,1)r+\omega_{3}(\nu,3)\frac{r}{6}(r+1)(r-4)\right]=\sum_{\nu}e^{-\beta
E(\nu)},
\end{equation}
we get
\begin{equation}
\omega_{3}(\nu,1)r+\omega_{3}(\nu,3)\frac{r}{6}(r+1)(r-4) = 1.
\end{equation}
Furthermore, the substitution of Eqs. (56), (57) and (58) into Eq.
(33) gives
\begin{equation}
\sum_{\nu}e^{-\beta
 E(\nu)}\left[-\frac{G}{4}(N-\nu)(r-N-\nu+2)+\omega_{3}(\nu,1)
G\frac{N(N-1)r(r-2)}{12}\right] = 0.
\end{equation}
This leads to
\begin{equation}
-\frac{G}{4}(N-\nu)(r-N-\nu+2)+\omega_{3}(\nu,1)G\frac{N(N-1)r(r-2)}{12}
= 0.
\end{equation}
Hence
\begin{equation}
\omega_{3}(\nu,1)=\frac{3(N-\nu)(r-N-\nu+2)}{N(N-1)r(r-2)},
\end{equation}
and
\begin{equation}
\omega_{3}(\nu,3)=\frac{6}{r(r-4)(r+1)}\left[1-\frac{3(N-\nu)(r-N-\nu+2)}{N(N-1)(r-2)}\right].
\end{equation}

Finally, we obtain the first-order, the second-order, and the
third-order reduced density matrices of $N$-particle systems
governed by the pairing Hamiltonian at any finite temperature as
shown below,
\begin{eqnarray}
 D^{1}(\beta)& = &\frac{1}{r}\sum_{i=1}^{r}|i><i|Z(\beta,V,N), \\
\nonumber  D^{2}(\beta) &=& \sum_{\nu}e^{-\beta E(\nu)} \frac{(N-\nu)(r-N-\nu+2)}{N(N-1)r}D^{2}(0)+ \\
&&\nonumber +\sum_{\nu}e^{-\beta E(\nu)}
  \left\{\frac{4(N-\nu)(r-N-\nu+2)}{N(N-1)r(r-2)^{2}}+
  \frac{2}{(r-2)(r+1)}\right.\\
&&\times\left.\left[1-\frac{3(N-\nu)(r-N-\nu+2)}{N(N-1)(r-2)}\right]\right\}D^{2}(2),
\end{eqnarray}
and
\begin{eqnarray}
\nonumber D^{3}(\beta) &=& \sum_{\nu}e^{-\beta E(\nu)}
\frac{3(N-\nu)(r-N-\nu+2)}{N(N-1)r(r-2)}D^{3}(1)+\sum_{\nu}e^{-\beta
E(\nu)}
\frac{6}{r(r-4)(r+1)} \\
&&\times\left[1-\frac{3(N-\nu)(r-N-\nu+2)}{N(N-1)(r-2)}\right]D^{3}(3).
\end{eqnarray}
These density matrices decide the microscopic states of the
system, from which we can calculate the thermodynamic properties.

\vspace{0.35in}

\section{Conclusions}

We have derived the second quantized form of the reduced Bloch
equations for both the canonical and grand canonical ensembles.
The deduction is based on the occupation number representation of
the contraction operator, and is similar to the ones we used
before as described in papers [1] and [2]. The equations obtained
provide a direct route for $\it{exactly}$ solving the reduced
density matrices for the systems represented by the model
Hamiltonian in the occupation number representation. They thereby
open an avenue for investigating the microscopic structures of
these systems and their relations to the equilibrium thermodynamic
properties.

As an example for exactly solving the reduced Bloch equations, we
have obtained the reduced density matrices of the pairing
Hamiltonian with different orders as shown in Eqs. (66), (67) and
(68). One very interesting conclusion we can make is that these
reduced density matrices can yield the projection of the ground
state of a superconductor in the $N$-particle space $V^{N}$. This
is the case when $G<0$ and $N$ is an even number, and the ground
state of the paring Hamiltonian (22) is an $AEGP$ (Antisymmetrized
Extreme Geminal Power) function $|N,(0^{\lambda}),(0^{\lambda})>$.
Therefore, if we choose $E(0)$ as the energy zero point and let
$\nu = 0$ and $\beta\rightarrow\infty$ in Eqs. (66), (67) and
(68), we can obtain the reduced density matrices of the ground
state for the superconductor as follows
 \begin{eqnarray}
\nonumber  D^{1}(T=0)&=&\frac{1}{r}\sum_{i=1}^{r}|i><i|
=\frac{1}{r}I_{1}, \\
 \nonumber D^{2}(T=0)&=&\frac{r-N+2}{r(N-1)}D^{2}(0)+\frac{2(N-2)}{r(r-2)(N-1)} D^{2}(2) \\
 \nonumber
 &=&\frac{r-N}{(r-2)(N-1)}|g><g|+\frac{2r(N-2)}{(r-2)(N-1)}D^{1}\wedge
 D^{1},
\end{eqnarray}
 and
 \begin{eqnarray}
 \nonumber D^{3}(T=0)&=&\frac{3(r-N+2)}{(N-1)r(r-2)}D^{3}(1)+\frac{6(N-4)}{r(r-2)(r-4)(N-1)}D^{3}(3) \\
 \nonumber &=&\frac{9(r-N)}{r(r-2)(r-4)(N-1)}|g><g|\wedge D^{1}+\\
 && +\frac{6r^{2}(N-4)}{(r-2)(r-4)(N-1)}D^{1}\wedge D^{1}\wedge
 D^{1}.
\end{eqnarray}
 These results are identical to Coleman's [33].

Since there are many other interesting physical systems often
modelled in an approximate way by the Hamiltonians with a second
quantized form, we expect that the ideas and methodologies we have
discussed in this paper can be further explored, and that the
fundamental equations (15) or (19) we have derived  will find wide
applications in the study of their microscopic states as well as
their interplay with macroscopic thermodynamic properties.

\vspace{0.35in}
\appendix
\section{Appendix: Second Quantized Form of Contraction and Expansion Operators}

Consider an antisymmetrized $N$-particle state ket
$|i_{1}i_{2}...i_{N}>$ composed of the single-particle states
defined in the Section 2. It is easy to see that the action of an
annihilation operator $a_{j}$ on this state ket is
\begin{equation}
a_{j}|i_{1}i_{2}...i_{N}> =
\sum_{k=1}^{N}(-1)^{N-k}\delta_{ji_{k}}|i_{1}i_{2}...\hat{i}_{k}...i_{N}>,
\end{equation}
where the superscript $\hat{ }$ indicates that the state $|i_{k}>$
is destroyed. The conjugate of Eq. (70) is
\begin{equation}
<i_{1}i_{2}...i_{N}|a^{+}_{j}=
\sum_{k=1}^{N}(-1)^{N-k}\delta_{ji_{k}}<i_{1}i_{2}...\hat{i}_{k}...i_{N}|.
\end{equation}
The combination of Eqs. (70) and (71) therefore yields
\begin{eqnarray}
\nonumber &&\sum_{i}a_{i}|i_{1}i_{2}...i_{N}><j_{1}j_{2}...j_{N}|a^{+}_{i}  \\
&=&\sum_{k,k^{'}}(-1)^{k+k^{'}}\delta_{i_{k}j_{k^{'}}}
|i_{1}i_{2}...\hat{i}_{k}...i_{N}><j_{1}j_{2}...\hat{j}_{k^{'}}...j_{N}|.
\end{eqnarray}

On the other hand, the state ket $|i_{1}i_{2}...i_{N}>$ can be
expanded as a determinant as follows,
\begin{equation}
|i_{1}i_{2}...i_{N}>=\frac{1}{\sqrt{N}}\sum_{k=1}^{N}(-1)^{k-1}
|i_{1}i_{2}...\hat{i}_{k}...i_{N}>|i_{k}>_{(N)}.
\end{equation}
Thus, the action of contraction operator $L_{N}^{N-1}$ on the
basis operator $|i_{1}i_{2}...i_{N}><j_{1}j_{2}...j_{N}|$ is
\begin{eqnarray}
\nonumber &&L_{N}^{N-1}\left( |i_{1}i_{2}...i_{N}><j_{1}j_{2}...j_{N}|\right)  \\
&=&\frac{1}{N}\sum_{k,k^{'}}(-1)^{k+k^{'}}\delta_{i_{k}j_{k^{'}}}
|i_{1}i_{2}...\hat{i}_{k}...i_{N}><j_{1}j_{2}...\hat{j}_{k^{'}}...j_{N}|.
\end{eqnarray}
The comparison of Eqs. (71) and (74) leads to
\begin{equation}
L_{N}^{N-1}\left(|i_{1}i_{2}...i_{N}><j_{1}j_{2}...j_{N}|\right)=
\frac{1}{N}\sum_{i}a_{i}|i_{1}i_{2}...i_{N}><j_{1}j_{2}...j_{N}|a^{+}_{i}.
\end{equation}
Furthermore, since any $N$th-order tensor operator $T^{N}$ can be
expressed as a linear combination of the basis operators
$\{|i_{1}i_{2}...i_{N}><j_{1}j_{2}...j_{N}|\}$, we finally get
\begin{equation}
  L^{N-1}_{N}(T^{N}) = \frac{1}{N}\sum_{i} a_{i} T^{N} a^{+}_{i},
\end{equation}
which is the second quantized form for the contraction operator.

 The expansion operator $\Gamma^{N}_{p}$ is the reverse operation
 of the contraction operator $L^{p}_{N}$ and transforms a
 $p$th-order tensor to an $N$th-order one $(p\le N)$ [17]
 \begin{equation}
  \Gamma^{N}_{p} (T^{p})= A_{N}\left[ T^{p}\otimes I^{N-p}\right]
  A_{N},
\end{equation}
 where $A_{N}$ is the antisymmetrized operator for $N$ particles.
 Since the expansion operator $\Gamma^{N}_{p}$ is the adjoint mapping of the
 contraction operators $L^{p}_{N}$ [17]
 \begin{equation}
  <\Gamma^{N}_{p} (T^{p}), T^{N}> = <T^{p},L^{p}_{N}(T^{N})>,
\end{equation}
 we can also obtain its second quantized form as follows
  \begin{equation}
  \Gamma^{N}_{N-1}(T^{N-1}) = \frac{1}{N}\sum_{i} a^{+}_{i} T^{N-1} a_{i}.
\end{equation}


\vspace{0.45in}

\begin{thebibliography}{99}
\vspace{0.15in}
\bibitem{wei1} L. Wei and C. C. Sun, Physica A, in press (2003);
e-print, cond-mat/0306306.
\bibitem{wei2} L. Wei and C. C. Sun, Physica A, in press (2003);
e-print, cond-mat/0306307.
\bibitem{bloch}F. Bloch, Zeits. f. Physick 74, 295 (1932).
\bibitem{kirkwood} J. G. Kirkwood, Phys. Rev. 44, 31 (1933).
\bibitem{friend} K. Pichler, D. A. Halliday, D. D. C. Bradley, P.
L. Burn, R. H. Friend, and A. B. Holmes, J. Phys.: Condens. Matter
5, 7155 (1993).
\bibitem{klemperer} W. Klemperer, Ann. Rev. Phys. Chem. 46, 1
(1995).
\bibitem{lin} J. Yu, M. Hayashi, S. H. Lin, K.-K. Liang, J. H. Hsu, W.
S. Fann, C.-I. Chao, K.-R. Chuang, S.-A. Chen, Synth. Met. 82, 159
(1996).
\bibitem{deleuze} S. P. Kwasniewski, J. P. Francois, and M. S.
Deleuze, J. Phys. Chem. A 107, 5168 (2003).
\bibitem{drickermer} H. G. Drickamer and C. W. Franck, Electronic
Transitions and the High Pressure Chemistry and Physics of Solids
(London, Chapman and Hall, New York, 1973).
\bibitem{kohn} W. Kohn and L. J. Sham, Phys. Rev. A 140, 1133
(1965).
 \bibitem{lorzao} (a) B. Lorzao, Nucl. Phys. A397, 225 (1983); (b)
 B. Lorzao and C. Quesne, Nucl. Phys. A403, 327 (1983).
  \bibitem{hagino} K. Hagino and G. F. Bertsch, Nucl. Phys. A679,
  163 (2000).
 \bibitem{blatt} J. M. Blatt, Prog. Theor. Phys. 23, 447 (1960).
 \bibitem{balian} R. Balian, H. Flocard and M. Veneroni, Phys.
 Rep. 317, 252 (1999).
 \bibitem{goscinski} B. Weiner and O. Goscinski, Phys. Rev.
 A 27, 57 (1983).
  \bibitem{ohrn} B. Weiner, H. J. Aa. Jensen and Y. $\ddot{O}$hrn, J. Chem.
  Phys. 80, 2009 (1984).
\bibitem{kummer} H. Kummer, J. Math. Phys. 8, 2063 (1967).
\bibitem{harriman1} J. E. Harriman, Phys. Rev. A 17, 1257 (1978).
\bibitem{cohen} L. Cohen and C. Frishberg, Phys. Rev. A 13, 927
(1976).
\bibitem{nakatsuji} H. Nakatsuji, Phys. Rev. A 14, 41 (1976).
\bibitem{schlosser} H. Schlosser, Phys. Rev. A 15, 1349 (1977).
\bibitem{harriman2} J. E. Harriman, Phys. Rev. A 19, 1893 (1979).
 \bibitem{valdemoro} C. Valdemoro, Phys. Rev. A 45, 4462 (1992).
 \bibitem{mazziotti} D. A. Mazziotti, Phys. Rev. A 57, 4219 (1998).
 \bibitem{coleman1} A. J. Coleman, Intern. J. Quantum Chem. 85,
 196 (2001).
\bibitem{coleman2} Z. H. Zeng, C. C. Sun, and A. J. Coleman, in
Density Matrices and Density Functionals, pp. 141-165 (1987), eds
by R. Erdahl and V. H. Smith Jr. (Dordrecht: Reidel).
\bibitem{nilsson} S. G. Nilsson and O. Prior, Mat.
Fys. Medd. Dan. Vid. Selsk. 32, no 16 (1960).
\bibitem{richardson2} R. W. Richardson, J. Math. Phys. 18, 1802 (1977).
\bibitem{draayer} F. Pan and J. P. Draayer, Phys. Lett. B 442, 7
(1998).
 \bibitem{wybourne} B. G. Wybourne, Classical Groups for
 Physicists (Wiley, New York, 1974).
\bibitem{sun}Z. H. Zeng and C. C. Sun, J. Mol. Sci. 4, 287 (1986).
 \bibitem{coleman3} A. J. Coleman, J. Math. Phys. 27, 1933 (1986).
 \bibitem{coleman4} A. J. Coleman, J. Math. Phys. 6, 1425 (1965).

\end{thebibliography}
\end{document}